\documentclass{PoS}

\newcommand{\ie}{i.e.}

\title{A new look at the cosmic ray positron fraction}

\ShortTitle{Cosmic ray positrons}

\author{Mathieu Boudaud \\
LAPTh, Universit\'e Savoie Mont Blanc \& CNRS, 9 Chemin de Bellevue, B.P.110 Annecy-le-Vieux, F-74941, France\\
E-mail: \email{mathieu.boudaud@lapth.cnrs.fr}}
     
\abstract{The positron fraction in cosmic rays was found to be steadily increasing in function of energy, above $\sim$10 GeV. This behaviour contradicts standard astrophysical mechanisms, in which positrons are secondary particles, produced in the interactions of primary cosmic rays during the propagation in the interstellar medium. The observed anomaly in the positron fraction triggered a lot of excitement, as it could be interpreted as an indirect signature of the presence of dark matter species in the Galaxy. Alternatively, it could be produced by nearby astrophysical sources, such as pulsars. Both hypotheses are probed in this work in light of the latest AMS-02 positron fraction measurements. The transport of primary and secondary positrons in the Galaxy is described using a semi-analytic two-zone model. MicrOMEGAs is used to model the positron flux generated by dark matter species.
We provide mass and annihilating cross section that best fit AMS-02 data for each single annihilating channel as well as for combinations of channels. We find that the mass of the favoured dark matter candidates is always larger than 500 GeV. The description of the positron fraction from astrophysical sources is based on the pulsar observations included in the ATNF catalogue. The region of the distance-to-age plane that best fits the positron fraction for a single source is determined and a list of five pulsars from the ATNF catalogue is given. Those results are obtained with the cosmic ray transport parameters that best fit the B/C ratio. Uncertainties in the propagation parameters turn out to be very significant.}

\FullConference{The 34th International Cosmic Ray Conference,\\
		30 July- 6 August, 2015\\
		The Hague, The Netherlands}

\begin{document}

\section{Introduction}

The cosmic ray positron flux at the Earth exhibits above 10~GeV an excess with respect to the astrophysical background produced by the interactions of high-energy protons and helium nuclei with the interstellar medium (ISM)~\cite{1997ApJ...482L.191B,2001ApJ...559..296D,2004PhRvL..93x1102B}. Recently, the Alpha Magnetic Spectrometer (AMS-02) collaboration has published \cite{Accardo:2014lma} an update on the positron fraction based on high statistics with measurements extending up to 500~GeV. The observed excess of positrons was readily interpreted as a hint of the presence of dark matter particles in the Milky Way halo. A  number of dark matter candidates have been proposed so far. The most favoured option is a weakly interacting massive particle (WIMP), whose existence is predicted by several theoretical extensions of the standard model of particle physics. Although marginal, WIMP annihilations are still going on today, especially in the haloes of galaxies where dark matter (DM) has collapsed, and where they produce various cosmic ray species. The hypothesis that the positron anomaly could be produced by the annihilation of DM particles is supported by the fact that the energy of the observed excess lies in the GeV to TeV range, where the WIMP mass is expected.
A completely different approach relies on the existence of pulsars in the Earth vicinity. These conventional astrophysical sources are known to release in the ISM electrons and positrons. Shortly after~\cite{2009Natur.458..607A} confirmed the positron anomaly, \cite{2009JCAP...01..025H} showed that the observations could easily be explained in this framework. Their conclusion was confirmed by \cite{2012CEJPh..10....1P} and recently, \cite{2013ApJ...772...18L} concluded that either Geminga or Monogem, two well-known nearby pulsars, could produce enough positrons to account for the AMS-02 precision measurements \cite{Aguilar:2013qda}.
In this work we reanalyse the cosmic ray positron excess in the light of the latest AMS-02 release \cite{Accardo:2014lma} and we thoroughly explore whether or not DM particles or a local conventional astrophysical source can account for this anomaly.


\section{Cosmic ray transport and the background}

Charged cosmic rays propagate through the magnetic fields of the Milky Way and are deflected by its irregularities. Cosmic ray transport can be modelled  as a diffusion process~\cite{2002PhRvD..65b3002C} and diffusion has been assumed to be homogeneous everywhere inside the Galactic magnetic halo (MH). The MH is modelled as a thick disc that matches the circular structure of the Milky Way. The Galactic disc of stars and gas, where primary cosmic rays are accelerated, lies in the middle. Positrons also lose energy as they diffuse. They spiral in the Galactic magnetic fields, emitting synchrotron radiation, and they undergo Compton scattering on the CMB and stellar light. The solution of the positron transport equation is obtained using Green functions~\cite{2009A&A...501..821D,2010A&A...524A..51D} as well as Bessel expansions~\cite{2008PhRvD..77f3527D}. 
The astrophysical background consists of the secondary positrons produced by the collisions of high-energy protons and helium nuclei on the atoms of the ISM. The production rate of secondary positrons can be safely calculated at the Earth and has been derived as in~\cite{2009A&A...501..821D}.
We then compute the total positron flux at the Earth $\Phi_{e^{+}} = \Phi_{e^{+}}^{\rm sec} + \Phi_{e^{+}}^{\rm prim}$, where the primary component is produced either by DM particles or by pulsars. The calculation is performed consistently with the same cosmic ray propagation model for both components. In most of this work, we have used the MED configuration, which best fits the boron to carbon ratio B/C~\cite{2004PhRvD..69f3501D}.
The positron fraction is defined as ${\rm PF} = {\Phi_{e^{+}}}/{\Phi_{L}}$, i.e. the ratio between the positron flux and the lepton flux $\Phi_{L} = \Phi_{e^{+}} + \Phi_{e^{-}}$. Usually, the electron flux is derived theoretically to get $\Phi_{L}$. However, contrary to positrons, the astrophysical background of electrons has a strong contribution, which is accelerated with nuclear species in supernova shock waves. This primary component is consequently very model dependent. We have therefore used new measurement of the lepton flux $\Phi_{L}$ from AMS-02~\cite{leptons_ICHEP_2014} to derive the positron fraction more accurately. Using the experimental lepton flux implies an additional error on the positron fraction arising from the error bars of lepton flux data. Therefore, we performed our analysis taking into account the total uncertainty, which is dubbed \textit{corrected errors} in the following figures.


\section{Dark matter analysis}
\label{sec:DM_analysis}

\begin{figure*}[ht!]
\begin{center}
\includegraphics[width=.49\textwidth]{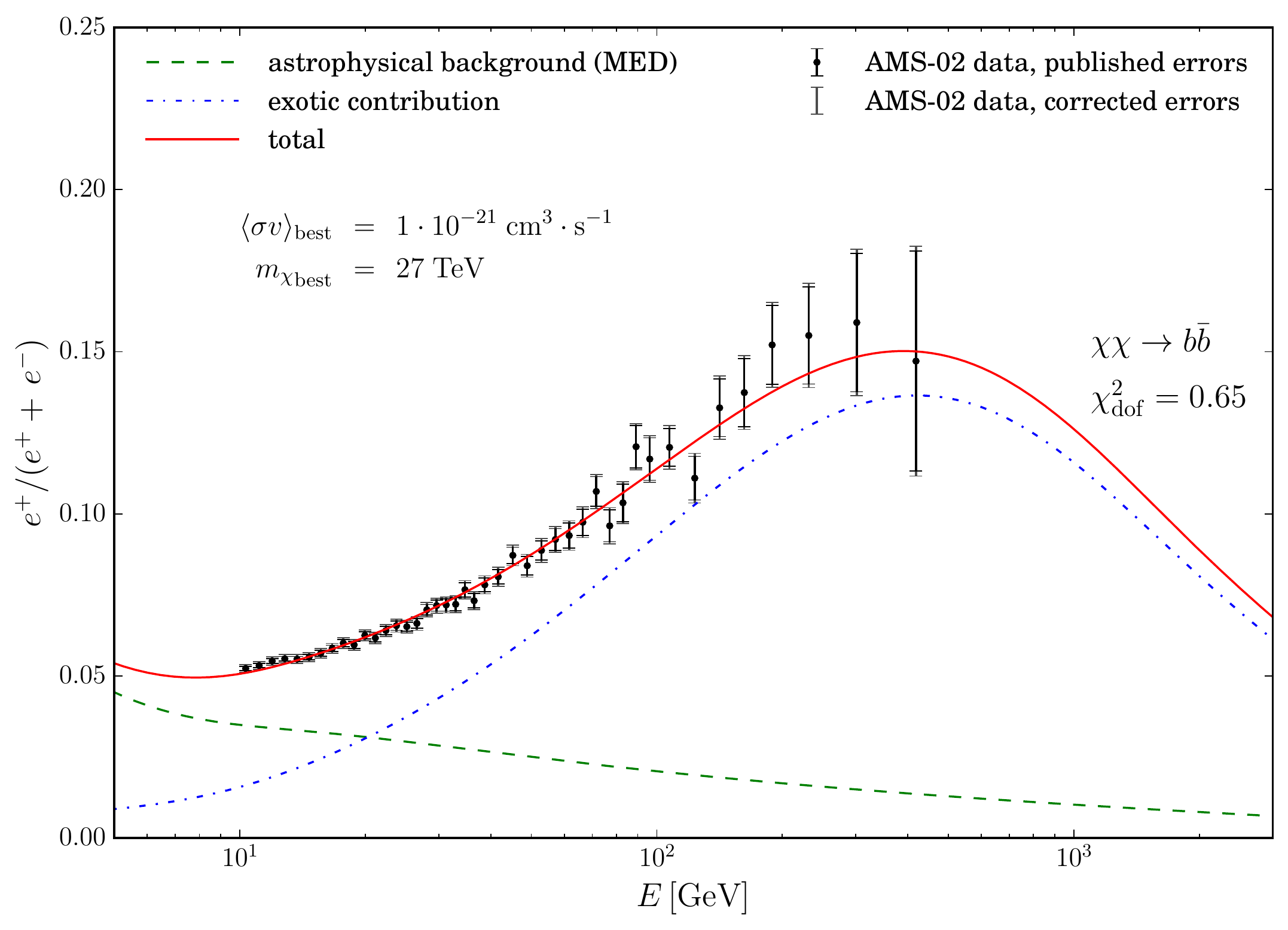}
\includegraphics[width=.49\textwidth]{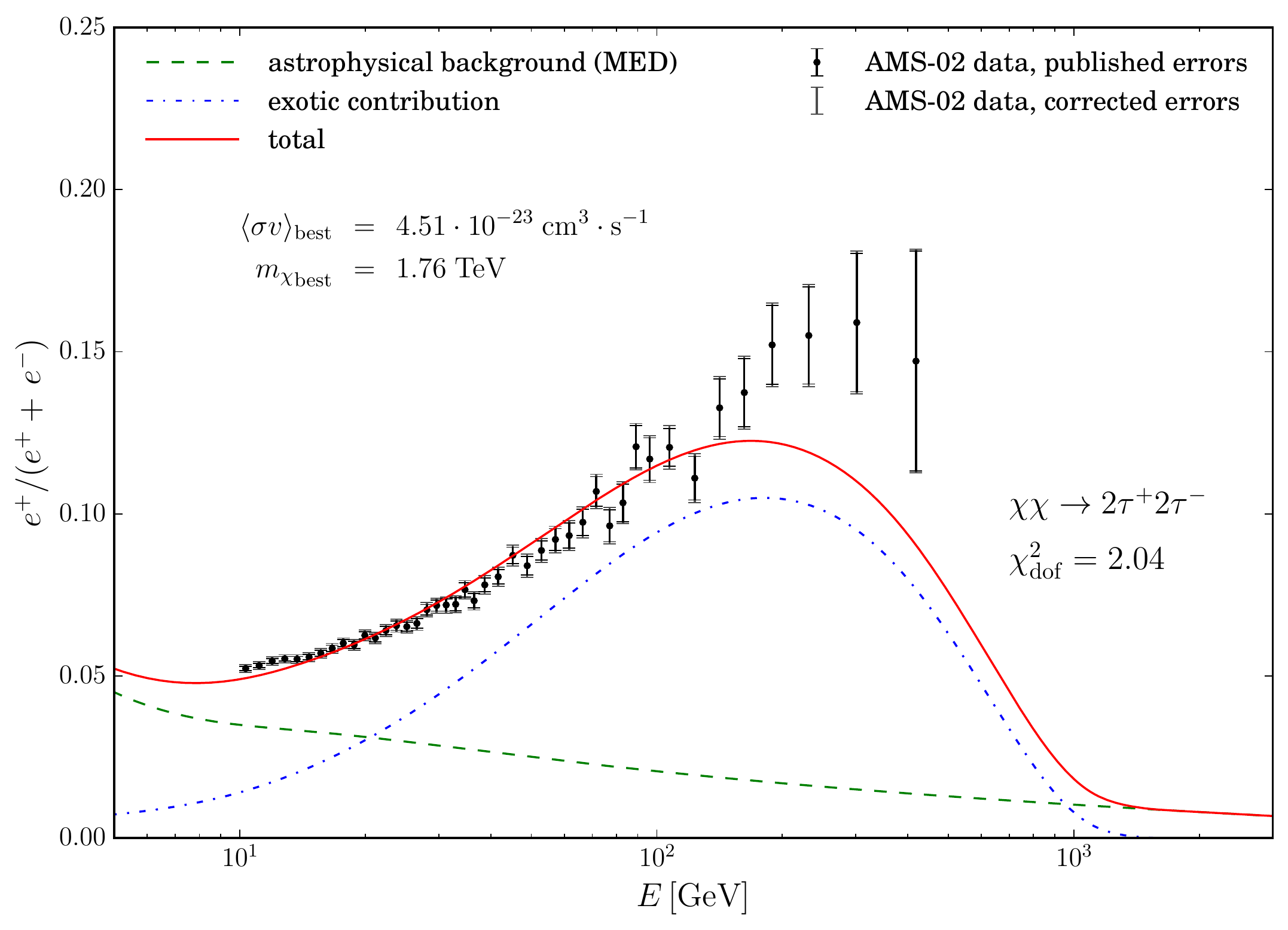}
\caption{Positron fraction as a function of the positron energy corresponding to the best-fit value of $\langle \sigma v \rangle$ and DM mass $m_{\chi}$ for $b \bar{b}$ (left) and $4\tau$ annihilation channels (right), compared with AMS data~\cite{Accardo:2014lma}. The propagation parameters correspond to the MED model. The AMS-02 lepton spectrum \cite{leptons_ICHEP_2014} is used to derive the $\chi^{2}$.}
\label{fig:fit_single_channel}
\end{center}
\end{figure*}

As a first interpretation of the AMS-02 results, we investigate the possibility that the excess of positrons at high energies originates from DM annihilation. The positron flux resulting from DM annihilation is computed with micrOMEGAs$\_{3.6}$~\cite{2011CoPhC.182..842B,2014CoPhC.185..960B}. Assuming a specific DM annihilation channel, we scan over two free parameters, the annihilation cross section $\langle \sigma v \rangle$ and the mass $m_{\chi}$ of the DM species. A fit to the AMS-02 measurements of the positron fraction is performed using MINUIT to determine the minimum value of the $\chi^{2}$.
To assess the goodness of our fit, we calculate the {\it p}-value from the $\chi^{2}_{n}$ test statistic with $n$ degrees of freedom obtained from each fit. We furthermore define two critical {\it p}-values for which we accept the resulting fit based on a 1 ($p > 0.3173$) and 2 ($p > 0.0455$) standard deviation ($\sigma$) significance level for a normal distribution.
Our results are presented in Table 1 of ~\cite{2015A&A...575A..67B}. We find that the data can be fitted very well with $p > 0.3173$ for annihilation channels into quark and boson final states whereas lepton final states provide very bad fits with $p < 0.0455$. The case where DM annihilates into four leptons, for example through the annihilation into a pair of new scalar (or vector) particles that decay into lepton pairs, does not provide better alternatives with $p < 0.0455$.
Figure~\ref{fig:fit_single_channel} shows best fit positron fraction spectra for the b-quark (left) and 4-$\tau$ (right) annihilation channels.

The description of DM annihilation into a single channel may be too simplistic.  Indeed, in most models annihilation proceeds through a combination of channels. Here we consider this possibility. To avoid introducing many free parameters and since the spectra are rather similar for different types of quarks for gauge and Higgs bosons, we only use the $b \bar{b}$ flux to describe quark final states. To a certain extent, spectra are also similar  for gauge and Higgs bosons since both decay dominantly into hadrons. For each case study, we use the fitting procedure described above, adding the branching fractions into specific channels as free parameters and scanning over the DM mass $m_{\chi}$.
As a first example, we consider the leptophilic case corresponding to the favoured DM candidate that originally explained the PAMELA positron excess without impacting the antiproton spectrum, as pointed out by \cite{2009NuPhB.813....1C,2009PhRvL.102g1301D}. We find a good fit, {\ie} with $\chi^{2}_{\rm dof}<1$, only for a DM mass near 500 GeV with a strong dominance of the $\tau^{+} \tau^{-}$ channel and only $10\%$ of direct annihilation into $e^{+}e^{-}$. This induces a sharper drop of the spectra near the last data point of AMS-02.
It is much easier to find excellent fits with $\chi^{2}_{\rm dof}<1$ when allowing for some hadronic channel and this for any DM mass in the range between 0.5 and 40~TeV.  The preferred branching fractions for the range of masses considered and the corresponding annihilation cross sections $\langle \sigma v \rangle$ are represented in Fig.~2 of~\cite{2015A&A...575A..67B}.
Finally allowing for any combination of the four-lepton channels allows for a very good fit to the data but only for a DM mass  between 0.5 and 1~TeV. Annihilation into $4\tau$ is by far dominant -- at least 70\%. Note that the $4e$ channel is subdominant and that the $4\mu$ channel is strongly disfavoured.

\begin{figure*}[ht!]
\begin{center}
\includegraphics[width=.49\textwidth]{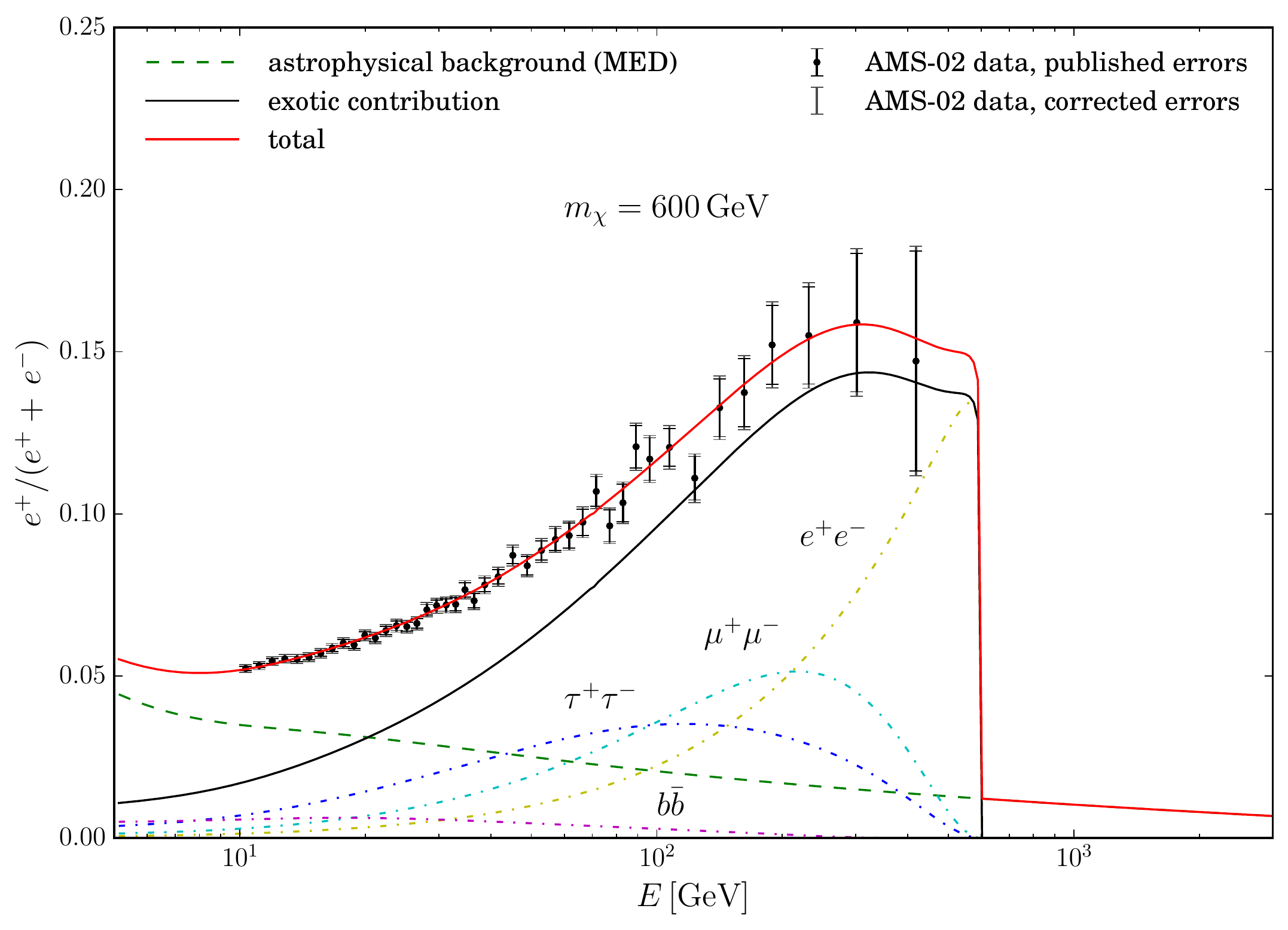}
\includegraphics[width=.49\textwidth]{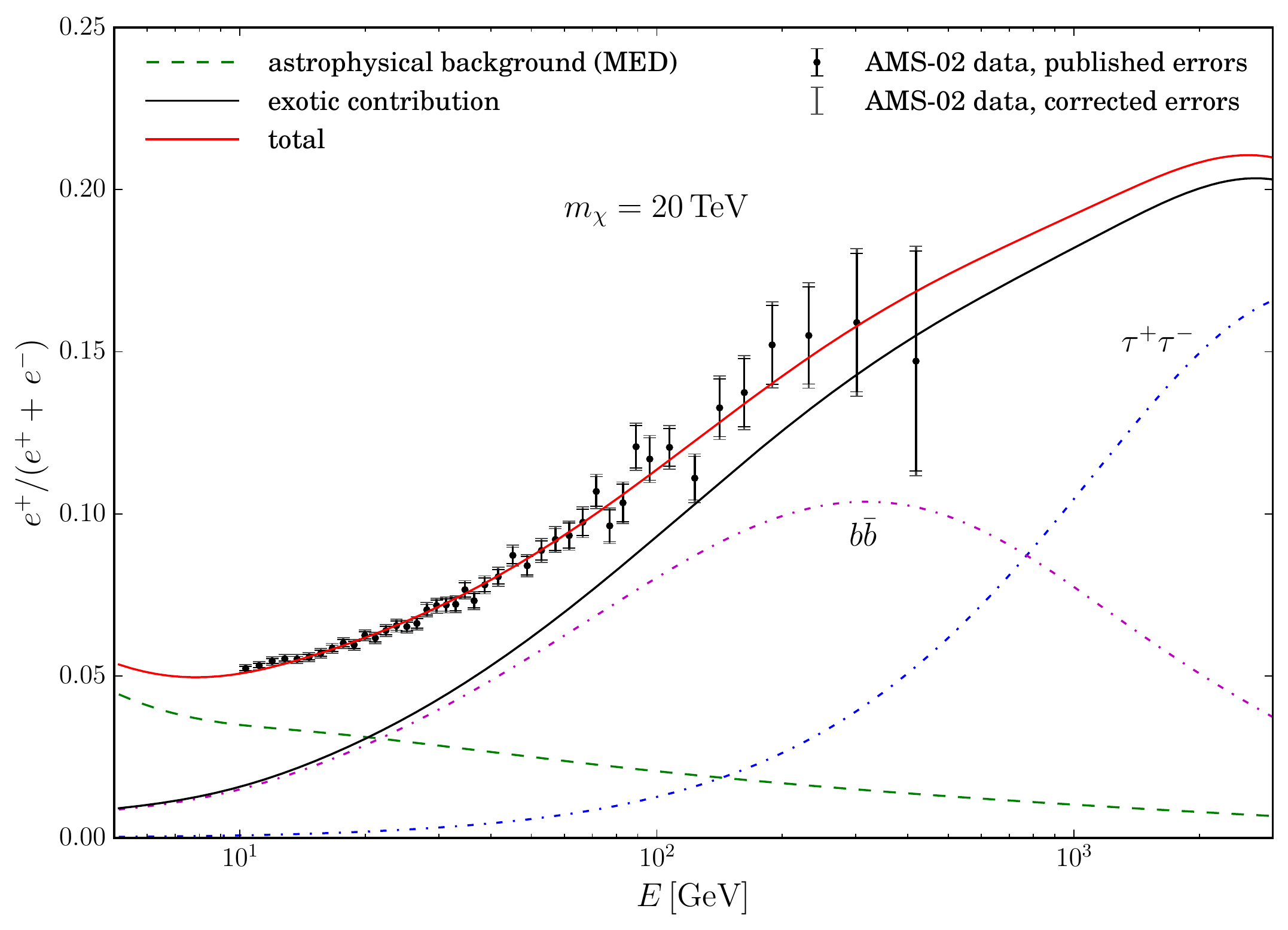}
\caption{Positron fraction as a function of the positron energy compared to AMS-02 data~\cite{Accardo:2014lma}. Left: m$_{\chi}$ = 600 GeV, $\langle \sigma v \rangle = 1.11 \cdot 10^{-23}$ cm$^{3}$ s$^{-1}$ and $\chi^{2}_{\rm dof} = 0.5$.
Right:  m$_{\chi}$ = 20 TeV, $\langle \sigma v \rangle = 1.09 \cdot 10^{-21}$ cm$^{3}$ s$^{-1}$ and $\chi^{2}_{\rm dof} = 0.6$. Corresponding branching ratios into lepton and $b \bar{b}$ pairs, are presented in Fig.~4 of~\cite{2015A&A...575A..67B}.
}
\label{fig:fit_combination_channel}
\end{center}
\end{figure*}


\section{The single pulsar hypothesis reinvestigated}

\begin{figure*}[h!]
\begin{center}
\includegraphics[height=0.40\textwidth]{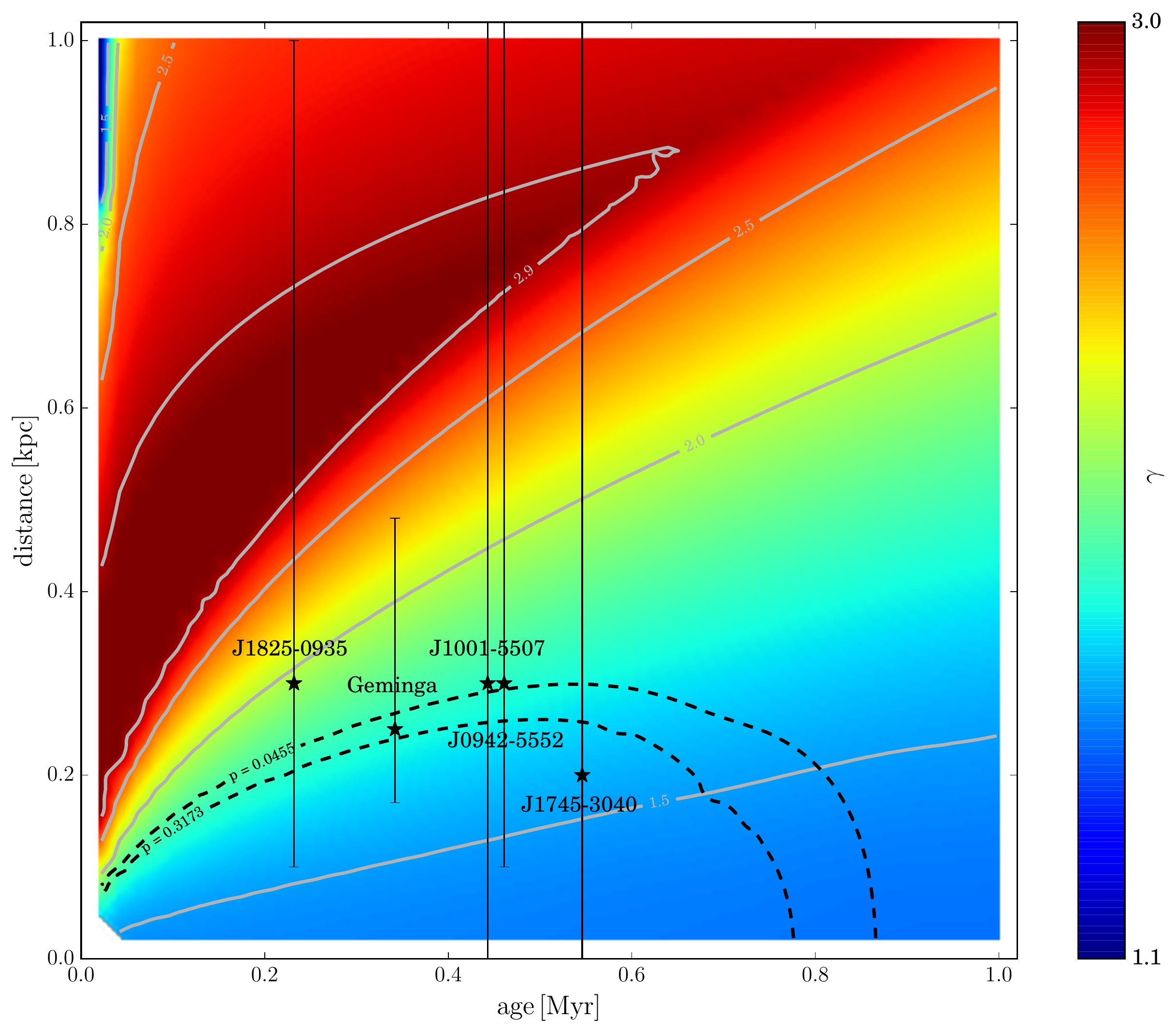}
\includegraphics[height=0.40\textwidth]{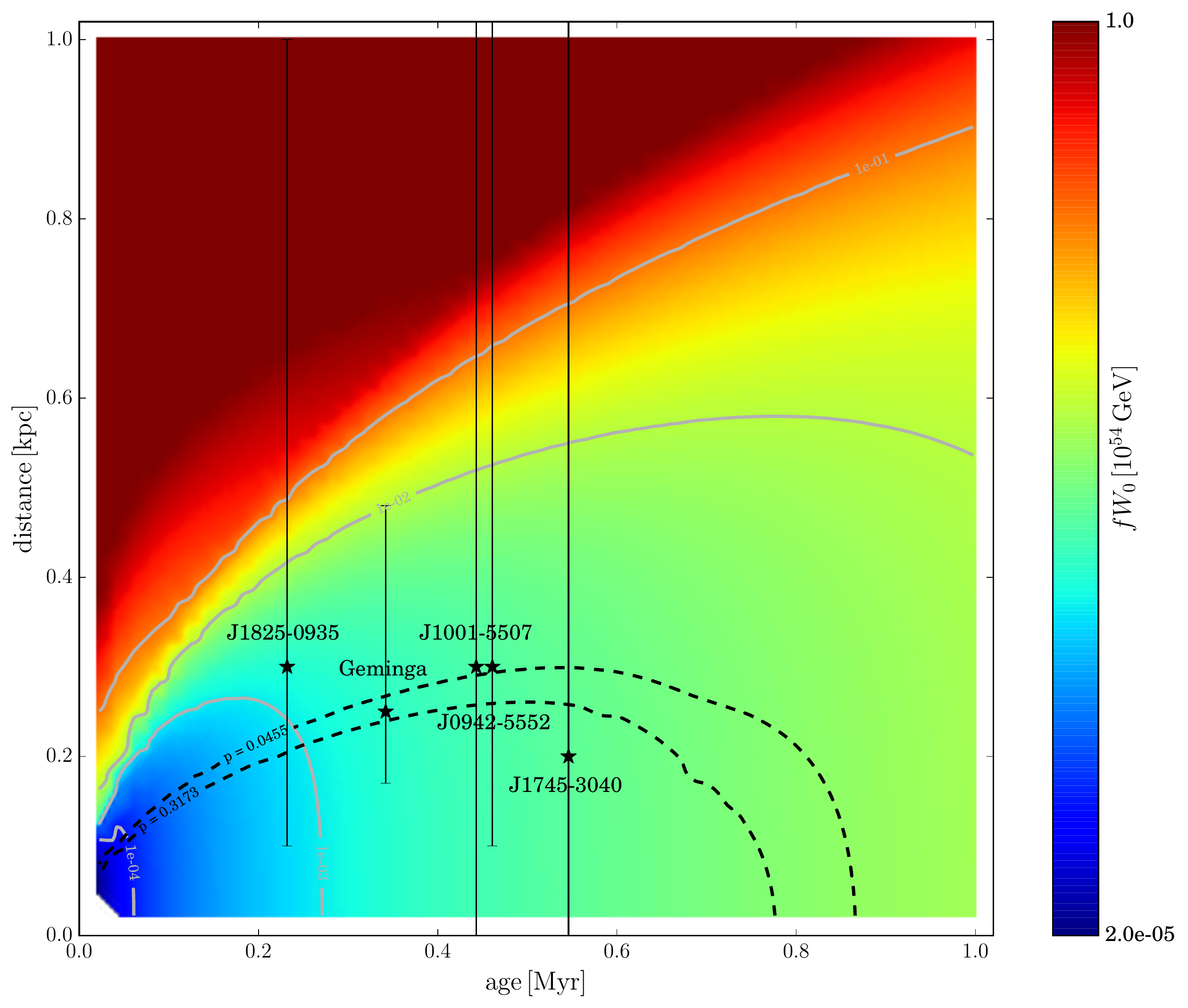}
\caption{Best-fit values of the spectral index $\gamma$ (left panel) and the total energy carried by positrons $fW_0$ (right panel) for each point of the plane (age, distance) with the benchmark propagation model MED. The grey lines display the iso-contours for given values of $\gamma$ (left) and $fW0$ (right). The black dashed lines represent the iso-contours of the critical {\it p}-values. The five selected pulsars with their associated uncertainty on their distance are indicated by the black stars.}
\label{fig:pulsar_gamma_fW_0}
\end{center}
\end{figure*}

Following~\cite{2013ApJ...772...18L} and~\cite{2013PhRvD..88b3013C}, we investigate if the rise of the positron fraction measured by AMS-02 can be explained by a single pulsar contribution.
Assuming a pulsar origin for the rise of the positron fraction leads to a cumulative contribution from all detected and yet undiscovered pulsars. Nevertheless, demonstrating that the positron fraction can be explained by a unique pulsar contribution, provides us with a valid alternative to the DM explanation of this anomaly. If the single pulsar hypothesis is viable, the whole of pulsars is capable of reproducing the experimental data. Indeed, as there is only an upper limit on the injection normalisation $fW_0$, adjusting $fW_0$ for each individual pulsar will result in even better fits when more pulsars are added.
The contribution of a single pulsar is calculated using the injection spectrum given in Sec.~2 of~\cite{2015A&A...575A..67B}. The free parameters are the spectral index $\gamma$ and  the energy released by the pulsar through positrons $fW_0$, which are related to the spectral shape and normalisation, respectively. In our analysis, we assume a fictional source placed at a distance \textit{d} from the Earth and of age $t_\star$. We then estimate the parameters $\gamma$ and $fW_0$, which give the best fit to the positron fraction. We allow the spectral index $\gamma$ to vary from 1 to 3 and we fix the upper limit of $fW_0$ to $10^{54} \; \rm{GeV}$. Since only close and relatively young single pulsars well reproduce the experimental data, we repeat this procedure for 2500 couples of $(d,\,t_{\star})$ with $d<1\,\rm{kpc}$ and $t_{\star}< 1\,\rm{Myr}$. 
The results are shown in Fig.~\ref{fig:pulsar_gamma_fW_0} where the colour scale indicates the value of  $\gamma$ (left panel) and $fW_0$ (right panel).
In the same figures, the two iso-contours of the critical {\it p}-values (black dashed lines) as defined in Sec.~\ref{sec:DM_analysis} are represented. Those define the good-fit region with $\gamma \lesssim 2 $ and $fW_0$ within the range of $[10^{49}, 10^{52}] \; \rm{GeV}$. We select the pulsars from the ATNF catalogue that fall into this good-fit region. The pulsar distance suffers from large uncertainties, which are taken into account for the pulsar selection. The uncertainty on the pulsar age is negligible due to a precise measurement of its spin and spin-down. Only five pulsars from the ATNF catalogue fulfil the goodness-of-fit criteria. The chosen pulsars and their distance uncertainties are indicated in Fig.~\ref{fig:pulsar_gamma_fW_0} by black stars with error bars.
For each of these five selected pulsars we estimate the values of $\gamma$ and $fW_0$ that best reproduce the experimental data. The results are listed in Table 3 of~\cite{2015A&A...575A..67B} with the corresponding $\chi^2$ and {\it p}-values.
As can be seen in Fig.~\ref{fig:pulsar_gamma_fW_0}, for their nominal distances, the pulsar J1745$-$3040 (J1825$-$0935) reproduces best (worst) the AMS-02 positron fraction. In contrast, Monogem and Vela cannot adjust the data. Because of their very young age, they are not able to contribute to the low-energy positron fraction between 10 and 50\,GeV where the error bars are the smallest.
%
%

\section{The effect of the cosmic ray propagation uncertainties}

\begin{figure*}[ht!]
\begin{center}
\includegraphics[height=0.35\textwidth]{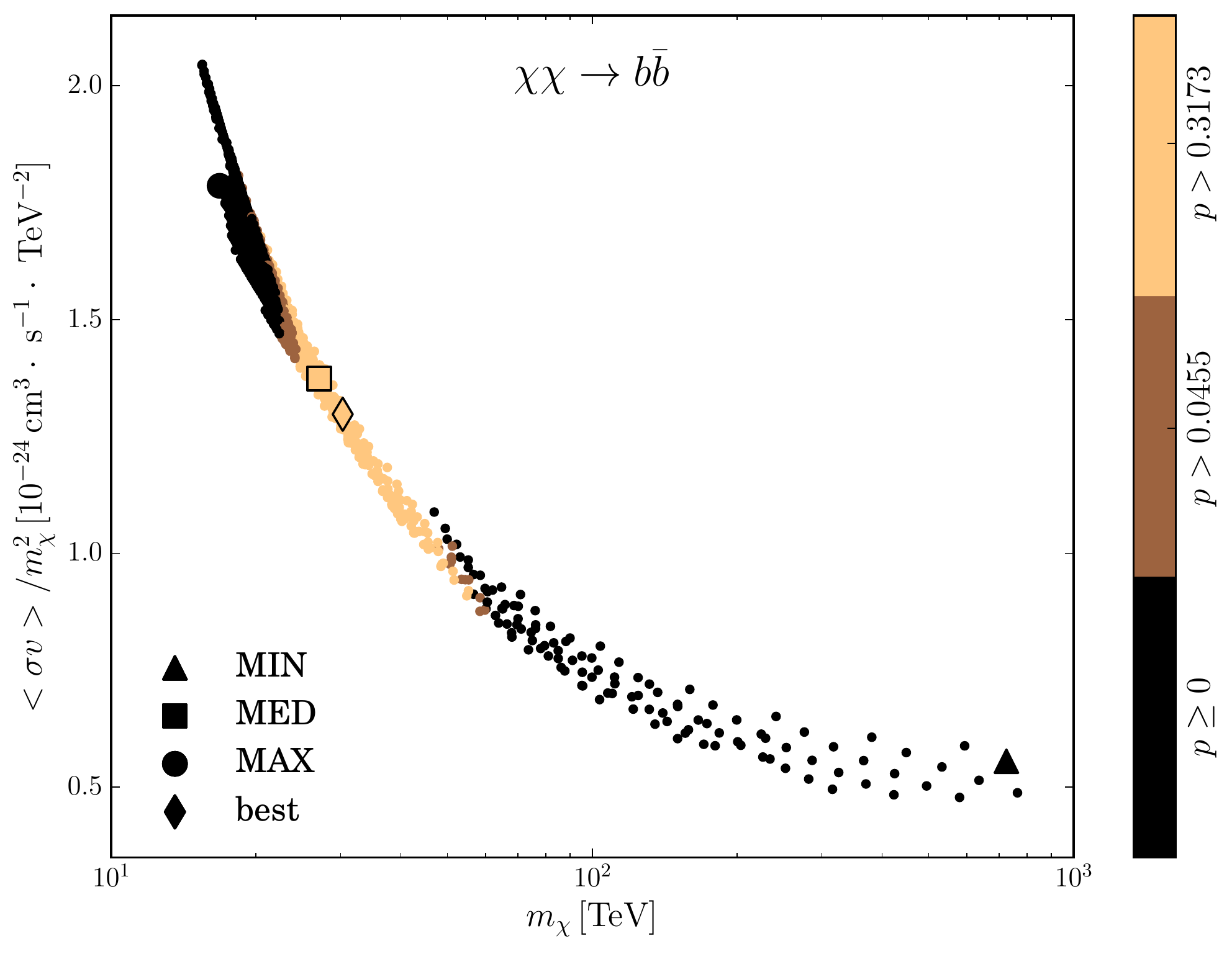}
\includegraphics[height=0.35\textwidth]{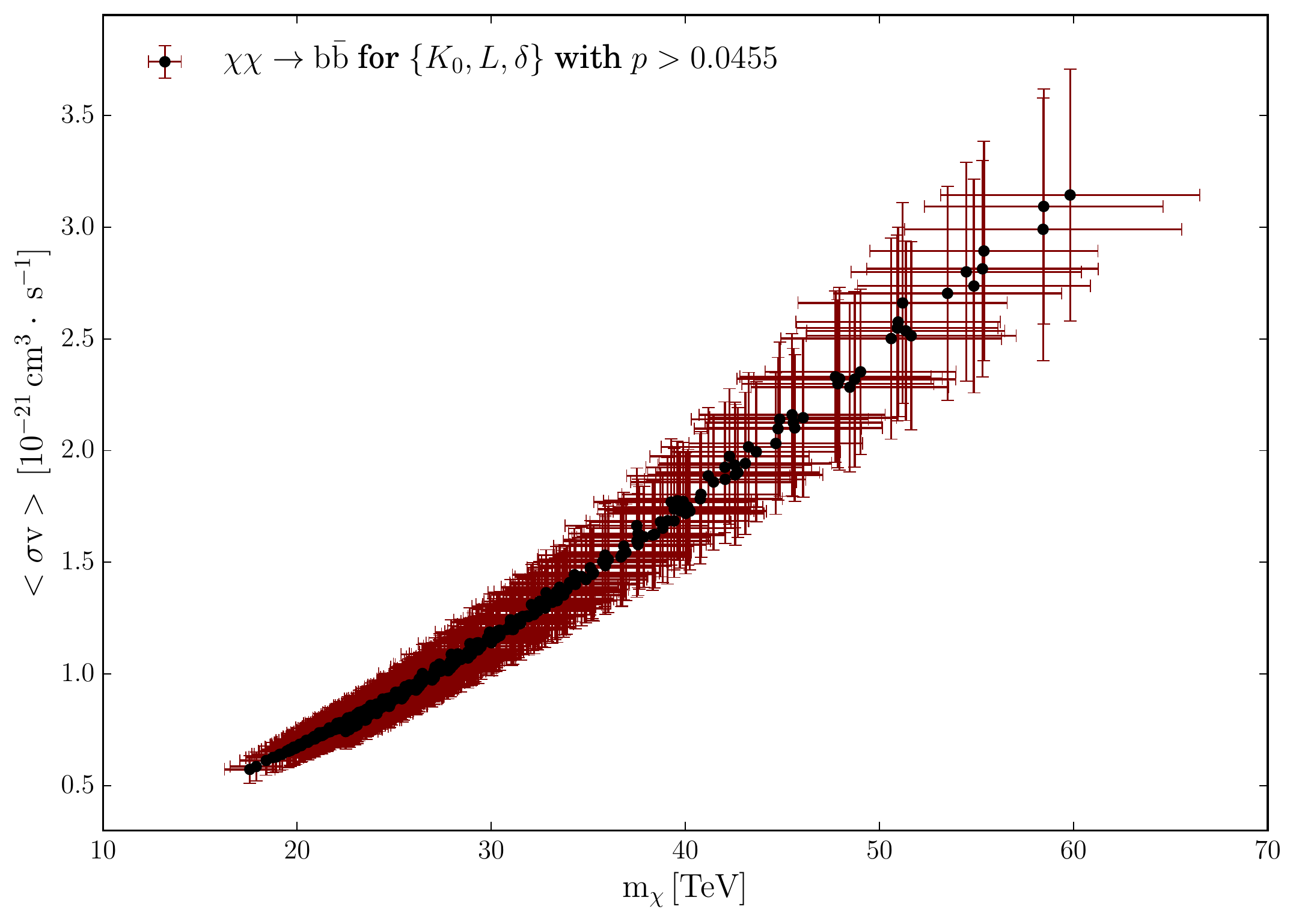}
\caption{ Left: {\it p}-value distributions of the 1623 transport parameter sets for the DM $\{m_\chi, \langle \sigma v \rangle/ m_\chi^2 \}$ parameters. The colour coding represents the increasing {\it p}-value from darker to lighter colours. The benchmark models MIN, MED, and MAX are represented with a triangle, square, and circle symbol, respectively. In addition, the best transport parameter set is highlighted with a diamond symbol. Right: Best-fit values of the DM $\{m_\chi, \langle \sigma v \rangle \}$ parameters for each transport parameter set with $p>0.0455$. The error bars represent the errors on the fit parameter resulting from the statistical uncertainty on the experimental data.}
\label{fig:propagation}
\end{center}
\end{figure*}

In this work we studied the constraints on an additional contribution of DM or a single pulsar to the positron fraction measured by the AMS-02 experiment above 10\,GeV. These constraints have been obtained by modelling the expected positron flux with the cosmic ray diffusion benchmark model MED defined in  \cite{2004PhRvD..69f3501D}. However, the transport mechanisms of Galactic cosmic rays are still poorly understood. The uncertainties on cosmic ray transport parameters are not negligible and have a major impact on searches for new physics. 
To take these uncertainties into account and to study their effect on modelling the positron fraction with an additional contribution, we use a set of 1623 combinations of the transport parameters $\{\delta, K_0, L, V_c, V_A \}$. 
These parameter sets result from the boron-to-carbon (B/C) ratio analysis by \cite{2001ApJ...555..585M}.
 In addition, the benchmark models MIN, MED, and MAX of~\cite{2004PhRvD..69f3501D}, widely used in the DM literature, are based on the parameters found in \cite{2001ApJ...555..585M}.
Note that the reaccelerating and convection processes are negligible for high energy positrons ($E>10 \rm \, GeV$) and are not taken into account in our positron flux calculation.
 We found a strong correlation between the transport and DM or pulsar parameters showing a huge impact on the best-fit values for the considered free parameters.
We only show in Fig.~\ref{fig:propagation} the $\chi \chi \rightarrow b\bar{b}$ channel as an example to highlight the correlations between the transport parameters and the parameters necessary to model the additional exotic contribution to the positron fraction at higher energies.

\section*{Conclusion}

This analysis aimed at testing the DM and pulsar explanations of the cosmic ray positron anomaly with the most recent available AMS-02 data. 
As regards the annihilating DM analysis, we found that AMS-02 data strongly disfavour a leptophilic DM. Nevertheless, the measurements are well explained when we allow annihilation in quarks and gauge or Higgs bosons, with a large annihilation cross section compare to the relic one ($\sim 3 \times 10^{-26}$ cm$^3$s$^{-1}$). 
These large DM annihilation rates, however, also yield gamma rays, antiprotons, and neutrinos for which no evidence has been found so far~\cite{2013arXiv1310.0828T,2012PhRvD..85f2001A,2014JCAP...02..008A,2014arXiv1410.2589H}. In particular, the upper limits on the DM annihilation cross section for a given mass and annihilation channel obtained from the observations of dwarf spheroidal galaxies challenge the DM interpretation of the positron anomaly. Recently, \cite{2015arXiv150101618L} have used our analysis combined with the Fermi/LAT dwarf galaxy data in order to rule out DM annihilation as an explanation of the  AMS-02 data.
In the same way, we have shown that the rise of the positron fraction can be alternatively explained by an additional contribution from a single pulsar. Indeed, five pulsars from the ATNF catalogue have been identified to satisfy the experimental measurements within their distance uncertainties. Demonstrating that the positron fraction can be explained by a unique pulsar contribution provides us with a valid alternative to the DM explanation of this anomaly.  As a matter of fact, if the single pulsar hypothesis is viable, the entirety of detected pulsars is hence capable of reproducing the experimental data.
The transport mechanisms of charged cosmic rays are still poorly understood, necessitating the inclusion of their uncertainties in the studies of the rise of the positron fraction. We observe that the error arising from the propagation uncertainties is much larger than the statistical uncertainty on the fitted parameters. In conclusion, the ignorance of the exact transport parameter values is the main limitation of such analyses. Henceforth, the study of cosmic ray propagation should be the main focus of future experiments.



\bibliographystyle{unsrt}

\bibliography{positrons}

\end{document}